\documentclass[preprint]{jpsj2}

\def\runtitle{Sound Propagation in Liquid ${}^3$He in Aerogel}
\def\runauthor{
Takayuki {\sc Ichikawa},
Mikio {\sc Yamamoto},
Seiji {\sc Higashitani} and
Katsuhiko {\sc Nagai}
}

\title
{
Collision Drag Effect on Propagation of Sound in
 Liquid ${}^3$He in Aerogel
}

\author
{ 
Takayuki {\sc Ichikawa},
Mikio {\sc Yamamoto},
Seiji {\sc Higashitani} and
Katsuhiko {\sc Nagai}
}

\inst
{
Faculty of Integrated Arts and Sciences,
Hiroshima University, 
Higashi-hiroshima, 739-8521
}

\recdate
{
\today
}

\abst
{
Sound propagation in a Fermi liquid with impurities is studied
using the Landau-Boltzmann equation and
the result is compared with a recent experiment in liquid ${}^3$He
in aerogel
by Northwestern University group. The sound absorption calculated
using the fixed impurity model is a few orders of magnitude
larger than the experiment. We take into account the simultaneous
motion of aerogel molecules and propose a model in which 
the momentum loss of ${}^3$He quasi-particles during the 
collisions with the aerogel
is converted to a drag force that acts on the aerogel
molecules. This collision drag model gives a reasonable
description for the temperature and the pressure dependence of
observed sound velocity and absorption.
}

\kword
{
Liquid ${}^3$He, Aerogel, Sound, Impurity Scattering,
Transport Relaxation Time, Collision Drag Effect, Fermi Liquid
}

\begin{document}
\sloppy
\maketitle

Liquid ${}^3$He in aerogel has recently attracted much attention as
a model system for studying impurity scattering effect in $p$-wave
pairing superfluid. When a homogeneous impurity model is assumed,
theoretical calculations are rather straightforward
\cite{Larkin, Volovik, Mineev, thuneberg, Rainer, Einzel, Kharadze,
Higashitani}.
The reduction of the transition temperature is
determined by the ratio of the coherence length $\xi$ to the so-called
transport mean free path $\ell_{\rm tr}$\cite{Larkin}. The  parameters in the
Ginzburg-Landau theory\cite{thuneberg} and
the superfluid density\cite{Higashitani} are written in terms of the
mean free path and the scattering cross section, when the ABM state or the
BW state is assumed. Sound propagation in dirty $p$-wave superfluid
has been also discussed\cite{Rainer, Maki}.

Since the discovery of superfluid transition of liquid ${}^3$He in aerogel
\cite{PoPa},
a lot of experiments  on superfluidity have been performed, using torsion oscillator\cite{MaPoPa, LKMP, PoPar},
 NMR\cite{Spa, Spb, Os, Bu, Hook},
acoustics\cite{Golov, Matsubara, Nomura, Gervais} and vibrating wire 
technique\cite{Pickett}. However even the  identification
of the superfluid phase  is not at satisfactory level.

Recently, Nomura et al.\cite{Nomura} and Gervais et al.\cite{Gervais}
reported high frequency 
acoustic experiment in the normal and the superfluid phase of liquid
${}^3$He in 98\% porous silica aerogel.  
The attenuation of sound is of the same order as in pure liquid ${}^3$He
but 
neither first-to-zero sound transition in the normal phase nor
collective mode attenuation peak which is characteristic to pure 
liquid ${}^3$He was observed.
These authors analyzed their data using the 
visco-elastic model.\cite{Halperin}
They suggested that the normal state attenuation can be obtained
by choosing the impurity scattering mean free path in the range of
200--300nm which is a bit larger than but of the same order 
as previously estimated values.

The aim of this letter is to discuss the sound propagation
taking into account the simultaneous motion of aerogel silica
molecules. The impurity scattering effect
cannot be fully treated by the visco-elastic model.
This model is essentially a modification of the
stress tensor such that can allow for the high frequency collisionless 
behavior\cite{Nagai}, but still assumes that the momentum density is conserved.
Since the impurity scattering does not conserve the momentum 
density\cite{Einzel},
the sound dispersion relation has a different form from that can be obtained
by the visco-elastic model. If the aerogel molecules are fixed, the sound
attenuation will become a few orders of magnitude larger than the observed
value. To remedy the situation we employ the idea of the
collision drag effect\cite{Pi, Ho}.
The momentum of ${}^3$He quasi-particles lost in the scattering
processes plays a role of drag force that acts on aerogel silica 
molecules. Since
the mass density of aerogel molecules is smaller than that of liquid
${}^3$He, the aerogel molecules move almost coupled with ${}^3$He
oscillation. The deviation in the motion between these two systems 
gives rise to a
damping of sound, in addition to the damping  in liquid
${}^3$He that can be accounted for by the visco-elastic model.

Let us consider the Landau-Boltzmann equation for the linearized distribution 
function $\delta f_{\vec{p}}e^{{\rm i}(\vec{q}\cdot\vec{r}-\omega t)}$ of the quasi-particle of liquid ${}^3$He in 
the normal phase
\begin{equation}
\omega\,\delta f_{\vec{p}}-\vec{v}_{\rm F}\cdot\vec{q}\left(\delta f_{\vec{p}}-f'\delta\epsilon_{\vec{p}}\right) = {\rm i}I[\delta f_{\vec{p}}],\label{LB}
\end{equation}
where $\omega, \vec{q}$ are the frequency and the wave vector of 
the disturbance, $\vec{v}_{\rm F}=\vec{p}_{\rm F}/m^*$ is the Fermi velocity, $f'$ is the
derivative of the equilibrium Fermi distribution function 
$f^0(\epsilon_{\vec{p}})$ with respect to the
quasi-particle energy $\epsilon_{\vec{p}}$ and $\delta\epsilon_{\vec{p}}$
is the additional quasi-particle energy due to the  Fermi 
liquid interaction. The right-hand-side is the collision integral that consists
of an impurity scattering term $I_{\rm imp}$ and a collision term due to the
mutual collisions between the quasi-particles.
\begin{equation}
I=I_{\rm imp}+I_{\rm coll}.
\end{equation}

In the longitudinal sound propagation, 
the angle dependence of $\delta f_{\vec{p}}$ is only through $\cos\theta
=\vec{p}\cdot\vec{q}/pq$. The solution of Eq.~(\ref{LB}) is known to have
a form
\begin{equation}
\delta f_{\vec{p}}=(-f')\sum_\ell y_{\ell} P_{\ell}(\cos\theta),
\end{equation}
where $P_\ell$ is the $\ell$-th Legendre polynomial. Using this notation,
Fermi liquid effect is given by
\begin{equation}
\delta\epsilon_{\vec{p}}=\sum_\ell \frac{F^S_{\ell}}{2\ell+1}y_{\ell} P_{\ell}(\cos\theta),
\end{equation}
where $F^S_{\ell}$'s are the symmetric part of the Landau parameters.

The impurity collision term due to the scattering by aerogel has a form
\begin{equation}
I_{\rm imp}=-\sum_{\vec{p}'}W_{\vec{p}\vec{p}'}2\pi\delta(\overline{\epsilon_{\vec{p}}}-\overline{\epsilon_{\vec{p}'}})(\delta f_{\vec{p}}-\delta f_{\vec{p}'}),
\end{equation}
where $W_{\vec{p}\vec{p}'}2\pi\delta(\overline{\epsilon_{\vec{p}}}-\overline{\epsilon_{\vec{p}'}})$ is the scattering probability and 
\begin{equation}
\overline{\epsilon_{\vec{p}}}=\epsilon_{\vec{p}}+\delta\epsilon_{\vec{p}}-\vec{p}\cdot \vec{v}.
\end{equation}
In the energy conservation, we have taken into account 
the Fermi liquid effect and also  the local velocity $\vec{v}$ 
of the aerogel. 
We assume that the aerogel motion is described by a
longitudinal displacement field
\begin{eqnarray}
\vec{u}(\vec{r},t)=
\vec{u}_{\vec{q}}e^{{\rm i}(\vec{q}\cdot\vec{r}-\omega t)},\qquad
\vec{u}_{\vec{q}}&=&u_q\,\vec{q}/q,
\end{eqnarray}
therefore, $\vec{v}=-{\rm i}\omega\vec{u}_{\vec{q}}$.

The local equilibrium is achieved when the ${}^3$He
quasi-particles move together with the scatterers, therefore the
local equilibrium distribution function $\delta f_{\rm l.e.}$
is given by
\begin{equation}
\delta f_{\rm l.e.}=f'\left(\delta \epsilon_{\vec{p}} -\vec{p}\cdot
(-{\rm i}\omega)\vec{u}_{\vec{q}}\right).\label{localeq}
\end{equation}
The collision term is a functional of
$\delta f'=\delta f_{\vec{p}}-\delta f_{\rm l.e.}$.
Bearing this into mind and defining the impurity 
scattering time $\tau_{\rm i}$
by
\begin{equation}
\frac{1}{\tau_{\rm i}}=\sum_{\vec{p}'}W_{\vec{p}\vec{p}'}2\pi\delta(\overline{\epsilon_{\vec{p}}}-\overline{\epsilon_{\vec{p}'}}),
\end{equation}
we find that $I_{\rm imp}$ is parametrized as follows:
\begin{equation}
I_{\rm imp}=\frac{-1}{\tau_{\rm i}}\left(\delta f' -
\sum_\ell (2\ell+1)\lambda^{\rm i}_{\ell}
\left\langle\delta f'P_{\ell}(\cos\theta)\right\rangle P_{\ell}(\cos\theta)
\right),\label{collimp}
\end{equation}
where $\langle\cdots\rangle$ stands for the angle average.
The number conservation at the impurity scattering requires that
\begin{equation}
\lambda^{\rm i}_0=1.
\end{equation}

Note that $\delta f_{\rm l.e.}$ of  Eq.~(\ref{localeq}) is also
a local equilibrium solution of the mutual collision term.
It follows that $I_{\rm coll}$ is quite satisfactorily approximated
by\cite{Peter}
\begin{equation}
I_{\rm coll}=\frac{-1}{\tau_{\rm c}}\left(\delta f' -
\sum_\ell (2\ell+1)\lambda^{\rm c}_{\ell}
\left\langle\delta f'P_{\ell}(\cos\theta)\right\rangle P_{\ell}(\cos\theta)
\right),\label{collcoll}
\end{equation}
where $\tau_{\rm c}$ is the mutual collision time which is well known to be
proportional to $1/T^2$. In the mutual collision process, the momentum
as well as the quasi-particle number is  conserved. Hence
\begin{equation}
\lambda^{\rm c}_0=\lambda^{\rm c}_1=1.
\end{equation}

Substituting these expressions into Eq.~(\ref{LB}), and summing up 
both the hand sides
over
$\vec{p}$ and the spin degrees of freedom, we find the number
 conservation law
\begin{equation}
\omega\,y_0=\frac{1}{3}v_{F}q\left(1+\frac{F_1^S}{3}\right)\,y_1.
\label{number}
\end{equation}
Multiplying $\cos\theta$ on both the hand sides of Eq.~(\ref{LB}) 
and summing up,
we obtain the momentum conservation law
\begin{eqnarray}
\omega\,y_1-v_{F}q\left((1+F_0^S)y_0+\frac{2}{5}\left(1+\frac{F^S_2}{5}\right)y_2\right)
=-\frac{{\rm i}}{\tau_{\rm tr}}
\left(\left(1+\frac{F^S_1}{3}\right)y_1+i\omega p_{\rm F} u_q\right),
\label{transport}
\end{eqnarray}
where $\tau_{\rm tr}$ is the so-called transport relaxation time defined by
\begin{equation}
\frac{1}{\tau_{\rm tr}}=\frac{1-\lambda^{\rm i}_1}{\tau_{\rm i}}.
\end{equation}
Since $F_0^S\gg 1$ in liquid ${}^3$He, we have only to estimate
$y_2/y_0$ up to order $1/F_0^S\sim(v_{\rm F}q/\omega)^2$. In the same way
as that in pure liquid ${}^3$He\cite{Peter}, we obtain
\begin{equation}
\frac{y_2}{y_0}=\frac{2\omega}{\displaystyle \omega + \frac{{\rm i}}{\tau_2}\left(1+
\frac{F_2^S}{5}\right)},
\end{equation}
where
\begin{equation}
\frac{1}{\tau_2}=\frac{1-\lambda^{\rm i}_2}{\tau_{\rm i}}+
\frac{1-\lambda^{\rm c}_2}{\tau_{\rm c}}.\label{visco}
\end{equation}
This $d$-wave contribution is that can be taken care of by the
visco-elastic model.

Let us first consider the case where the aerogel molecules are
fixed, i.e., $u_q=0$.
From Eqs.~(\ref{number}), (\ref{transport}) and (\ref{visco}),
we obtain the sound dispersion relation
\begin{eqnarray}
\omega^2+\frac{{\rm i}\omega}{\tau_{\rm{tr}}}\left(1+\frac{F_1^S}{3}\right)
=c_1^2q^2
\left(1+\frac{4}{5}\frac{1+\frac{F^S_2}{5}}{1+F_0^S}
\frac{\omega}{ \omega + \frac{{\rm i}}{\tau_2}\left(1+
\frac{F_2^S}{5}\right)}\right),\label{dispersion}
\end{eqnarray}
where $c_1$ is the first sound velocity of pure liquid ${}^3$He.
The damping term due to $\tau_{\rm{tr}}$ is characteristic to
the fixed impurity model and dominates the damping of sound.
In the {\it collisionless} regime 
$\omega\tau_{\rm{i}}, \omega\tau_{\rm{c}}\gg 1$, the
sound velocity is equal to the zero sound velocity $c_0$ of
pure ${}^3$He liquid, while the
absorption is given by
\begin{equation}
\alpha=\frac{1}{2c_0\tau_{\rm{tr}}}(1+\frac{F_1^S}{3}).
\end{equation}
The 
absorption is much larger than that in the pure liquid,
because the prefactor $(c_0^2-c_1^2)/c_1^2$ that appears in the
absorption of pure liquid is missing. 
In the {\it hydrodynamic} regime $\omega\tau_{\rm{i}}\ll 1$,
the sound dispersion becomes
\begin{equation}
\frac{{\rm i}\omega}{\tau_{\rm{tr}}}(1+\frac{F_1^S}{3})=c_1^2 q^2.
\end{equation}
The frequency used by Northwestern group\cite{Nomura, Gervais}
 is 15MHz.
If one takes the mean free path
$\ell\sim v_{\rm F}\tau_{\rm{tr}}\sim 270$nm, then $\omega\tau_{\rm{tr}}\sim 0.7$.
Thus the system is rather in the {\it hydrodynamic} regime and
the sound is hard to propagate in contrast to the experimental
observation. It is not possible to interpret the experiment
using the fixed impurity model.

Now we consider the effect of the motion of aerogel molecules.
We assume that in the skeleton aerogel the displacement
field $u(\vec{r},t)$ obeys the wave equation
\begin{equation}
\rho_a \ddot{u}= \rho_a c_a^2 \Delta u,
\end{equation}
where $\rho_a$ is the aerogel mass density and $c_a\sim 50$m/sec\cite{Golov,
Nomura, Gervais} is the sound velocity of the skeleton aerogel.
We can easily calculate from Eq.~(\ref{transport}) the 
momentum density transfer per unit time from ${}^3$He to
aerogel:
\begin{equation}
\frac{1}{\tau_{\rm tr}}\frac{N_{\rm F}p_{\rm F}}{3}\left(\left(1+\frac{F_1^S}{3}\right)y_1
+{\rm i}\omega p_{\rm F}u_q
\right),
\end{equation}
where $N_{\rm F}$ is the density of states of ${}^3$He quasi-particles
for both spin projections.
This is just the drag force density exerted to the aerogel molecules.
We have for the equation of motion of the aerogel
\begin{equation}
\rho_a(-\omega^2 + \omega_q^2)u_q=
\frac{1}{\tau_{\rm tr}}\frac{N_{\rm F}p_{\rm F}}{3}
\left(\left(1+\frac{F_1^S}{3}\right)y_1+{\rm i}\omega p_{\rm F}u_q\right),
\end{equation}
from which we find
\begin{equation}
u_q=\frac
{\displaystyle \frac{-1}{\tau_{\rm tr}}\frac{N_{\rm F}p_{\rm F}}{3}\left(1+\frac{F_1^S}{3}\right)y_1}
{\displaystyle \rho_a\left(\omega^2-\omega_q^2+{\rm i}\omega\frac{1}{\tau_{\rm tr}}
\frac{N_{\rm F}p_{\rm F}^2}{3\rho_a}\right)}.
\end{equation}
Substituting this into Eq.~(\ref{transport}), we find that the
dispersion relation of sound has the same form as 
Eq.(\ref{dispersion}), but $\tau_{\rm tr}$ is replaced by
$\tau_{\rm eff}$
$$
\frac{1}{\tau_{\rm eff}}=\frac{1}{\tau_{\rm tr}}
\frac{\omega^2-\omega_q^2}
{\displaystyle
\omega^2-\omega_q^2+\frac{{\rm i}\omega}{\tau_{\rm tr}}
\frac{N_{\rm F}p_{\rm F}^2}{3\rho_a}}.
$$

In the experiment by Northwestern group\cite{Nomura, Gervais},
 the observed sound velocity
is of order the first sound velocity. In this case, 
$\omega_q^2=c_a^2q^2$
can be neglected when compared with $\omega^2$. 
Hence $\tau_{\rm eff}$ is much simplified to
\begin{eqnarray}
\frac{1}{\tau_{\rm eff}}&=&\frac{1}{\tau_{\rm tr}}
\frac{\omega}
{\omega+\frac{{\rm i}}{\tau'}},\\
\frac{1}{\tau'}&=&
\frac{1}{\tau_{\rm tr}}\frac{\rho_3}{\rho_a}(1+\frac{F_1^S}{3}),
\end{eqnarray}
where we have used the fact that $N_{\rm F}p_{\rm F}^2$ is related to
the ${}^3$He mass density $\rho_3$ via
$
N_{\rm F}p_{\rm F}^2=3(1+F_1^S/{3})\rho_3
$.
The aerogel density $\rho_a$ is 0.04g/cm${}^3$\cite{Golov, Gervais}, 
while ${}^3$He
density is about 0.1g/cm${}^3$ at 15bar\cite{Wheatley}. 
Since $\omega\tau_{\rm tr}$
is small compared with unity, we can further approximate
\begin{equation}
\frac{1}{\tau_{\rm eff}}=-{\rm i}\omega\frac{\tau'}{\tau_{\rm tr}}
\left(1+{\rm i}\omega\tau'\right)=-{\rm i}\omega\frac{\rho_a}{\rho_3}
\frac{1+{\rm i}\omega\tau'}{1+\frac{F_1^S}{3}}.
\end{equation}

We finally find the dispersion relation
\begin{eqnarray}
\left(1+\frac{\rho_a}{\rho_3}\right)\omega^2
+{\rm i}\omega^3\left(\frac{\rho_a}{\rho_3}\right)^2
\frac{1}{1+\frac{F_1^S}{3}}\tau_{\rm tr}
=c_1^2q^2\left(1+\frac{4}{5}\frac{1+\frac{F^S_2}{5}}{1+F_0^S}
\frac{\omega}{\omega + \frac{{\rm i}}{\tau_2}\left(1+
\frac{F_2^S}{5}\right)}\right).\quad
\label{dispersionn}
\end{eqnarray}
Thus the sound velocity is
\begin{equation}
c^2=c_1^2/(1+\frac{\rho_a}{\rho_3})
\end{equation}
and the absorption is
\begin{equation}
\alpha=\frac{\omega^2}{c}
\left(\frac{1}{2}\frac{\displaystyle\left(\frac{\rho_a}{\rho_3}\right)^2}
{\displaystyle 1+\frac{\rho_a}{\rho_3}}\frac{\tau_{\rm tr}}{1+\frac{F_1^S}{3}}
+\frac{2}{5}\frac{\tau_2}{1+F_0^S}\right).
\end{equation}
The first term gives the absorption in the high temperature limit.
The second term explains the temperature dependence of $\alpha$
reported by Nomura et al.\cite{Nomura},
because
\begin{equation}
\tau_2=\left\{ \begin{array}{ll}
\tau_{\rm i}/(1-\lambda_2^{\rm i}) & \mbox{for} \quad T\rightarrow 0,\\
\propto 1/T^2\rightarrow 0 & \mbox{for}\quad T\rightarrow \infty.
\end{array}\right.
\end{equation}

The present result reproduces quite well the pressure dependence
of the sound velocity and the absorption reported by Northwestern
group\cite{Nomura, Gervais}.
To fit the reported value of the absorption $\alpha$,
however, we have to take $\ell_{\rm tr}=v_{\rm F} \tau_{\rm tr}$
or $\ell_{\rm i}= v_{\rm F}\tau_{\rm i}$
around 50nm, which is about 3 times smaller than the 
previous estimates\cite{Rainer, PoPar}. This discrepancy
is still to be examined. 

The collision drag effect is prominent when the condition
$\omega\tau_{\rm i}< 1$ is satisfied. As is evident from Eq.
(\ref{dispersionn}), the dragged aerogel gives an extra inertia
in the sound oscillation\cite{Mc}.
The low frequency experiment by Golov et al.\cite{Golov} in
the normal phase can be also interpreted by the present theory.
In the high frequency
limit $\omega\tau_{\rm i}\gg 1$, the aerogel molecules cannot move
and the result is the same as that of the fixed impurity
model. In this sense, the frequency 15MHz used by Northwestern
group is not a {\it high} frequency but a {\it low} frequency.

In conclusion, we have presented a microscopic theory of collision
drag effect on the sound propagation in liquid ${}^3$He aerogel system.
Extension of the theory to superfluid phase shall be reported
elsewhere.
 
We thank Dr.\ R. Nomura for discussions on the experimental results.
This work is supported in part by a Grant-in-Aid for Scientific Research
(No.11640354) and a Grant-in-Aid for COE
Research (No.13CE2002) of the Ministry of Education, Culture,
Sports, Science and Technology of Japan.

\end{document}